# Direct Observation of the Reciprocity between Spin Current and Phonon Interconversion

Swapnil Bhuktare,[1] Hanuman Singh,[1] Arnab Bose,[1] and Ashwin. A. Tulapurkar[1,*]

[1]*Department of Electrical Engineering, Indian Institute of Technology-Bombay, Powai, Mumbai 400076, India*

*Spin current has emerged as a leading candidate for manipulation of spins in a nano-magnet. We here experimentally show another utility of spin current viz. it can be used for generation of phonons. Within the same experimental setup, we also demonstrate the inverse effect of generation of spin current by phonons. To demonstrate them, we measured the scattering-matrix of a two-port device with interdigital transducers as one port and array of Ni/Pt lines as second port on piezoelectric substrate. The off-diagonal elements which correspond to transmission between the ports, were found to have $180^o$ relative phase shift. The transmission of electrical signal from port 2 to 1 corresponds to generation of phonons from spin-current, while transmission from port 1 to 2 corresponds to the inverse effect. These results could be useful for designing spin-current based gyrators.*

Spintronics exploits the spin degree of freedom of an electron for various new functionalities [1]. The coupling of spins to heat currents (spin caloritronics) [2-9] orbital momentum (spin orbitronics) [10-13] or the mechanical degrees of freedom (spin mechanics) [14-22] has generated a lot of interest due to the novel physics involved and potential new applications of spin current. This coupling has enabled observation of various new phenomena such as spin Seebeck effect [3-6], acoustically driven resonance [15-17] etc. Spin seebeck effect refers to the generation of spin current over a macroscopic scale by temperature gradient, whereas in acoustically driven resonance, mechanical motion is used to excite spin dynamics. We here demonstrate that spin current can be used to excite phonons, which can travel macroscopic distances of the order of millimeter. Our experiment utilizes a piezoelectric $LiNbO_3$ substrate over which we fabricate a periodic array of Ni/Pt lines. We observed that when current is passed through Ni/Pt lines, phonons in the form of surface acoustic waves (SAW) are emitted, which can be detected by using interdigital transducers (IDT). This phonon emission involves many spin based phenomena: i) Conversion of charge current into spin current by spin-Hall effect (SHE) in Pt  ii) Excitation of magnons in Ni by spin current via spin-transfer torque (STT) effect iii) emission of phonons by magnons via magneto-elastic coupling. Our setup also allows us to measure the inverse effect i.e. generation of spin current by phonons [19]. The inverse effect proceeds via inverse of the above three steps: i) generation of magnons by phonons via inverse magneto-elastic coupling ii) injection of spin current in Pt via.  inverse STT effect or spin pumping iii) Conversion of spin current into charge current by inverse spin-Hall effect (ISHE) in Pt.

Our experiments show that the transmission of electrical signal from IDTs to Ni/Pt lines ($S_{21}$) and transmission from Ni/Pt lines to IDTs ($S_{12}$) have opposite sign i.e. $S_{12}=-S_{21}$ i.e. the device shows non-reciprocal behavior. We however, found that the generalized reciprocity relation viz. $S_{12}(B)=S_{21}(-B)$, where B denotes the magnetic flux density, is always obeyed. As discussed in the above paragraph, the transmission of electrical signal between the two ports involves three processes. We studied the reciprocity between these three processes and their inverse effects separately. The reciprocity between SHE and ISHE is demonstrated in supplementary information by a local measurement on a 5 terminal device made with Pt cross and Ni pads. To demonstrate the reciprocity between STT and spin-pumping, we fabricated a two port device where the signal transmission takes place via combination of SHE and STT, and reverse transmission via combination of spin pumping and ISHE. The reciprocity is demonstrated by measuring the scattering matrix of the device. By fabricating suitable two port devices based on these effects we here present a 'direct' experimental proof of the generalized reciprocity by measuring the scattering matrices. It should be noted that though the STT and spin pumping effects have been experimentally demonstrated, the reciprocity between these effects has not been demonstrated directly. A direct experimental proof amounts to showing that the magnitudes of the two effects are equal. Futher, a thought experiment based such two port devices can be used to derive the expression for the inverse effect if the equation for direct effect is known by imposing the condition that $S_{ij}(B)=S_{ji}(-B)$.

**Experiments:**

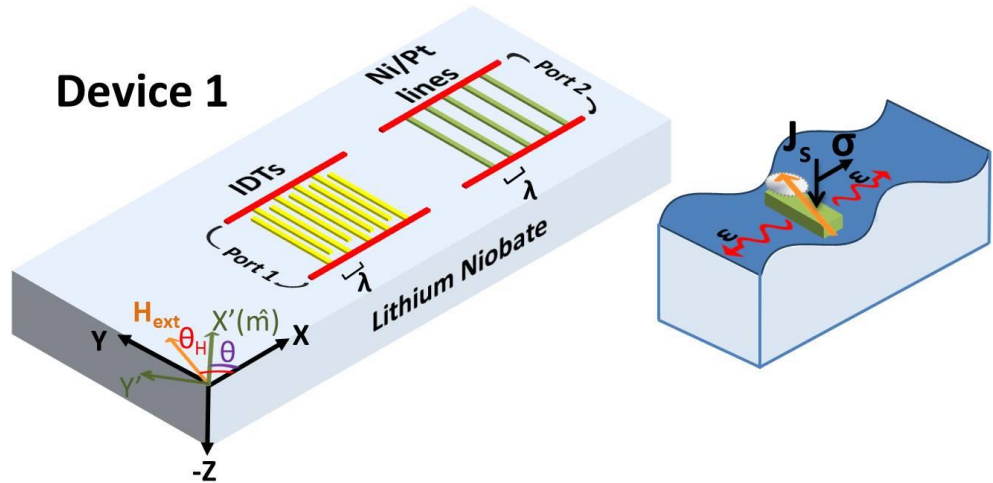

*Fig.1 The schematic of the experiment: Shorted Ni/Pt lines with periodic spacing are made on one side as port 2 and IDTs are made on another side as port 1. When RF charge current is passed through Pt, it converts it into spin current by virtue of the spin Hall effect. This spin current excites the magnetization of the Ni, which in turn emits phonons because of the magneto elastic coupling with the substrate. This process is shown schematically in the right hand panel. The phonons emitted by the Ni lines in the form of SAW interfere constructively and can then be detected at the other end by the IDTs. On the other hand, when the IDTs are excited, they emit*

*SAW which produces the acoustic spin pumping in the Ni/Pt lines and a voltage can be picked up at port 2.*

The device schematic is shown in Fig. 1. The fabrication of the device was carried out with the help of standard lithography followed by deposition and lift off techniques. We fabricated an array of 50 Ni/Pt lines with 2 µm separation. Each line is 300 µm long and 400 nm wide with thickness of Ni/Pt as 8 nm/6 nm. For IDTs, the length and width dimensions are the same as that of Ni/Pt lines and are composed of 10 nm Ti/ 50 nm Au. The distance between the centres of IDTs and Ni/Pt lines is about 1.5 mm. Before carrying out the S-parameters measurements, we performed the Spin torque FMR (ST-FMR) measurements [23,24] on Ni/Pt lines (Port 2) to ensure that we are able to excite Ni by the spin current generated by Pt. The dc voltage spectrum as a function of the magnetic field applied at 45º angle obtained by passing rf current with 4 GHz frequency is shown in Fig. 2(a). The spectrum after subtraction of constant background can be fitted to a combination of symmetric ($V_s$) and antisymmetric ($V_A$) Lorentzian functions given by

$$V_S = C_1 \frac{\Delta^2}{4(H-H_r)^2 + \Delta^2} \quad \text{and} \quad V_A = C_2 \frac{4(H-H_r)\Delta}{4(H-H_r)^2 + \Delta^2}$$

where $\Delta$ is the linewidth, $H_r$ is the resonance field. Note that $V_s$ shows the contribution of spin transfer torque excitation arising from the Spin Hall effect in Pt and $V_A$ shows the contribution of Oersted field excitation. The fit gave $C_1$=2.6 µV and $C_2$=2.8 µV implying we have a substantial spin torque excitation of the ferromagnet. The dc voltage measurement results for different values of the frequency are shown in Fig. 2(b). The resonant frequency versus magnetic field obtained from the dc voltage spectra is shown by the black dots in the inset of Fig. 2(b). The resonance frequency versus magnetic field follows Kittel's relation given below:

$$f_0 \approx (\gamma/2\pi)\sqrt{(H'_0 + H'_{//})(H'_{ext} + H'_{//} + H'_\perp)} \quad ---(1)$$

where, $H_0' = H_0 \cos(\theta - \theta_H)$, $H_{//}' = H_{//}(\cos^2\theta - \sin^2\theta)$, $H_\perp' = H_\perp + H_{//}\sin^2\theta$

where *$H_\perp$ and $H_{//}$* are out-of-plane and in-plane anisotropy fields respectively, $\theta_H$ is the angle between x –axis and magnetic field and, $\theta$ is the angle between x-axis and magnetization. The red curve in inset of Fig. 2b was obtained with the parameters, $H_{//}$=55 Oe, $H_\perp$ =4.2 kOe, $\gamma$=2.1×10$^5$ m/(A s).

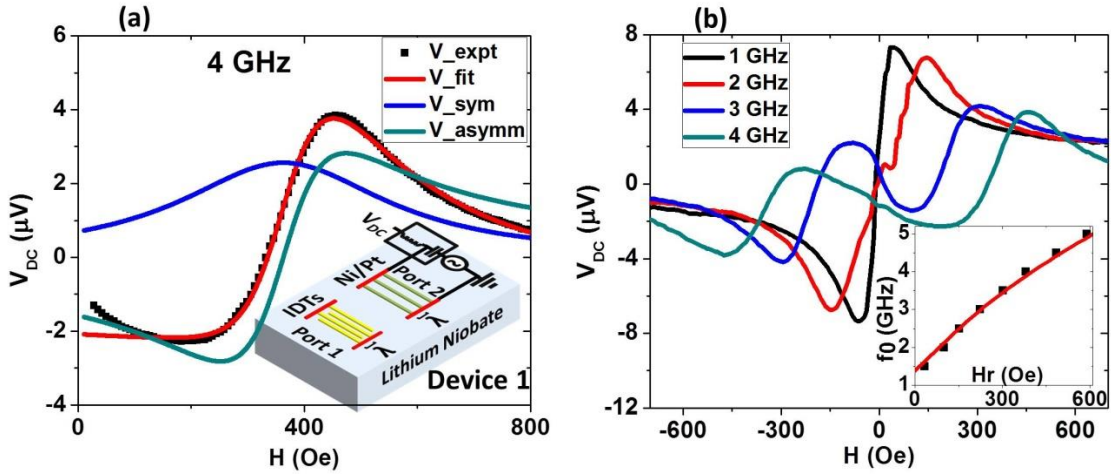

*Fig.2 ST-FMR study: (a) The dc voltage spectrum as a function of magnetic field applied at 45º angle for rf current of 4 GHz frequency. The black dots show the experimental data, the fitted red curve is a combination of symmetric Lorentzian (blue curve) and anti-symmetric Lorentzian (cyan curve) (b) The DC voltage spectra measured for different frequencies. The inset shows the Kittel's plot of resonance frequency versus magnetic field*

Next we carried out the scattering parameter measurements on the device by fixing the frequency to 1.92 GHz with 5 dBm power. (This frequency corresponds to the resonance of the IDTs which was experimentally determined from the dip in $S_{11}$ spectrum shown in the supplementary information [25].) The real and imaginary parts of the signal $S_{12}$ as a function of magnetic field applied at 45º angle are shown in Fig. 3(a). The real part shows a clear peak (or a dip) and the imaginary part shows the dispersion around 100 Oe magnetic field, which indicates signal transmission from port 2 to port 1 as the Ni undergoes FMR. This signal transmission is ascribed to generation of surface acoustic waves created by oscillating magnetization of Ni. As the Ni lines are periodically spaced with a spacing of λ, the emitted waves interfere constructively to give a large $S_{12}$ signal. When voltage is applied to port 2, the excitation of Ni is substantially by the spin current produced in Pt by spin-Hall effect. Thus the $S_{12}$ signal corresponds to interconversion of spin current into phonons. On the other hand, when voltage is applied to IDTs, they emit SAW, which produce an effective RF magnetic field on the Ni lines because of the magneto elastic coupling between the Ni and the piezoelectric substrate [15]. This magnetic field excites the magnetization of Ni which in turn pumps spin current into Pt. Pt converts this spin current into charge current via ISHE and thus a voltage appears across port 2 which is detected as $S_{21}$ signal as shown in Fig.3(b). Thus $S_{21}$ signal corresponds to the interconversion of phonons into spin current. It should be noted that the transmission of signal between the two ports involves a large delay (distance between the ports/saw velocity) which affects the phase of the $S_{12}$ and $S_{21}$ signals. Thus the shape of the real and imaginary parts can be a combination of peak and dispersion shapes depending on the delay [22]. From Fig 3(a) and 3(b), we see that $S_{12}$ and $S_{21}$ have 180 degree phase shift, however the generalized reciprocity relation $S_{12}(B)=S_{21}(-B)$ is obeyed. The above measurements were done for different angles of the applied magnetic field. No clear signals were observed for $\theta_H=0^0$ or $\theta_H=90^0$. The amplitude of the signal as a function of θ is shown in Fig. 3(c). It follows a $sin^2(\theta)cos(\theta)$ dependence and shows a four fold symmetry, a peculiar characteristic of the magneto elastic coupling [15, 22]. The peak position of $|S_{12}|$ as a function of θ is shown in Fig. 3(d). The simulated blue curve in Fig. 3(d) is obtained from this relation $S_{12} \alpha \chi_{11} sin^2(\theta)cos(\theta)$ [22].

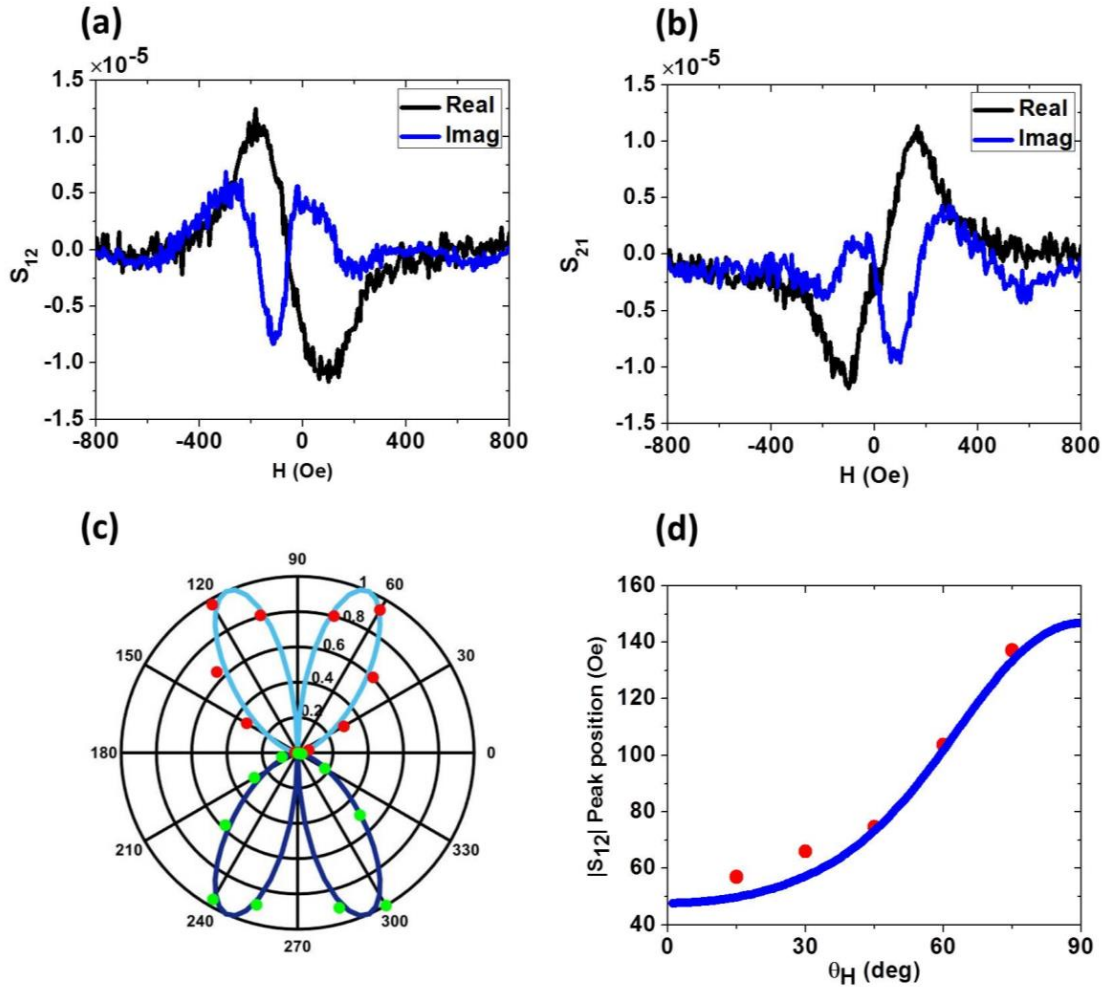

*Fig.3 S-parameters of device 1: the magnetic field was applied at an angle of $45^0$ for part (a) and (b) a) $S_{12}$ signal, Real and imaginary parts. b) $S_{21}$ signal, Real and imaginary parts. c) Variation of the amplitude of $S_{12}$ signal for magnetic field applied at different angles. The experimental data is shown by the red and green points and the $sin^2(\theta)cos(\theta)$ curve is shown by the continuous light and dark blue lines. b) Variation of the $|S_{12}|$ peak position. The experimental data is shown by the red points and the simulated values are shown by the continuous blue curve*

The generalized reciprocity between transmission in the above device, involves magneto-elastic coupling/inverse magneto-elastic coupling, spin transfer torque/spin pumping and spin-Hall effect/inverse spin-Hall effect. Now we show the reciprocity between spin pumping + inverse spin-Hall effect and spin transfer torque + spin-Hall effect. The schematic of the experiment is shown in Fig. 4(a) and (d). We made two devices on LiNbO$_3$ substrate: Device 2, with Ni/Pt (2 nm/6 nm) strip and Device 3, with Ni/Au (20 nm/80 nm) strip. A coplanar waveguide (CPW) electrically insulated by SiO$_2$ layer was fabricated on the top. The device fabrication was done with standard lithography (optical), deposition (sputtering) and lift-off techniques. The strip is connected to port 2 and the waveguide is connected to port 1. For Device 2, when port 2 is excited, Pt converts the RF charge current into spin current by the SHE, which excites the magnetization of Ni by the STT mechanism. The inductive coupling with the waveguide gives rise to a voltage at port 1, which is detected as $S_{12}$ signal. On the other hand, when port 1 is excited, the oscillating RF field produced by the waveguide excites Ni magnetization, which pumps spin current in Pt. Pt converts this spin current into charge current by the ISHE and a voltage appears across port 2, which is detected as $S_{21}$ signal. The $S_{12}$ and $S_{21}$ signals are shown in Fig. 4(b) and Fig. 4(c) respectively. We see that $S_{12}(B)=S_{21}(B)=S_{21}(-B)$ which implies that the combination of SHE and STT or ISHE and spin pumping are reciprocal to each other. We however do not have a gyrator behavior as observed for device 1. The observed reciprocity in particular shows that the same spin mixing conductance parameter determines both the STT and spin-pumping phenomena [26, 27] (See more discussion in the supplementary information). When Pt is replaced by thick Au (device 3), the excitation of Ni is because of the RF Oersted field only. The $S_{12}$ and $S_{21}$ signals are shown in Fig. 4(e) and (f) respectively, show the reciprocity in this device which is based on Ampere law and inductive coupling. When we have a dominant spin current excitation of FM, the real part of transmission shows a dispersion and imaginary part shows a dip (or peak) but when we have a dominant Oersted field excitation, the behavior is interchanged i.e. real part shows a peak and imaginary part shows dispersion.

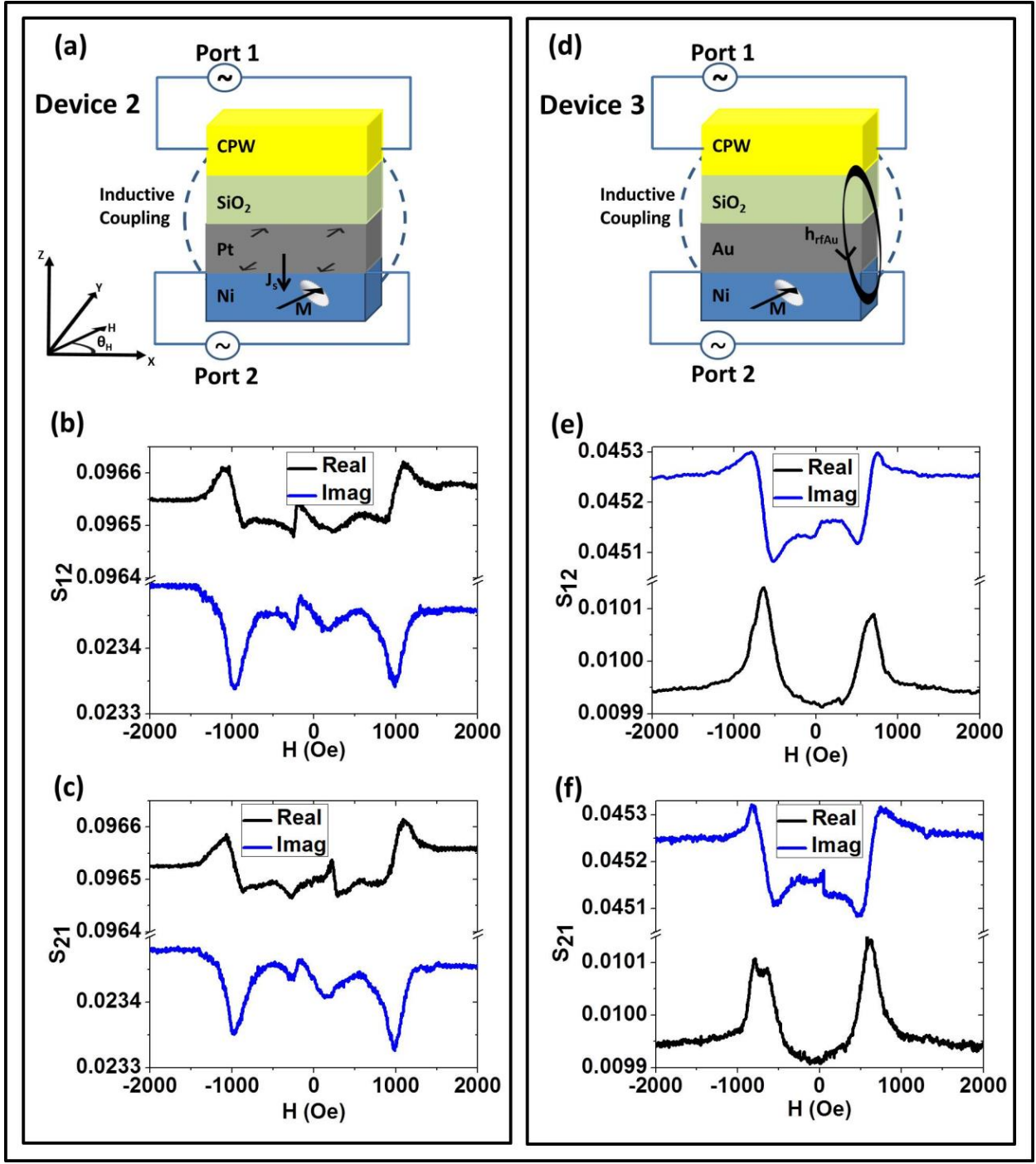

*Fig. 4 Left hand panel shows reciprocity between spin pumping and spin transfer torque. (a) The schematic of the device 2 and measurement setup. (b) and (c) $S_{12}$ and $S_{21}$ parameters of device 2 measured as a function of magnetic field. (e) and (f) panels show $S_{12}$ and $S_{21}$ for device 3 shown in panel d. The S-parameters satisfy the relation $S_{12}(B)=S_{21}(-B)$. All measurements were carried out at 5 GHz with 10 dBm power with magnetic field applied along $45^o$ angle.*

Finally, using a 5 terminal device we show that spin Hall and inverse spin Hall effects are also reciprocal to each other. The device schematic and measurement results are shown in the supplementary material.

In summary, we have demonstrated experimentally the spin current to phonon interconversion and its inverse process. We investigated different physical phenomena involved in this, and shown that each of these phenomena satisfies generalized reciprocity. Our results on spin current-phonon interconversion show that this effect can be exploited to design acoustic gyrators. This could provide new pathways for interconversion of spin currents and phonons and could be used for integrating spintronics with micro or nano-electromechanical systems and open new pathways into the emerging fields like spin mechatronics [28,29].

## Methods:

All the devices were fabricated on a $128^0$ Y cut lithium niobate substrates with a wave velocity of 3980 m/s. The patterning of the IDTs was done with e beam lithography with the help of a conducting polymer to avoid charging effects followed deposition of Cr/Au (10 nm/50 nm) (by thermal evaporation) and lift off. Ni/Pt (8 nm/6 nm) lines were then made using similar recipe. The big contact pads were then made by optical lithography followed by deposition of Cr/Au (10 nm/100 nm) and lift off. The measurements were carried with the help of Agilent N5244A network analyzer. Spin torque ferromagnetic resonance measurements were done on Port 2 to confirm the spin current excitation of Ni lines. The operational frequency was determined by the dip in the $S_{11}$. The frequency spectrums were recorded while sweeping magnetic field for different angles. Time gating and background subtraction were done to get the signals. All measurements were done at room temperature.

**Corresponding Author**

* Email: ashwin@ee.iitb.ac.in

**Author Contributions**

The device fabrication and measurements were carried out by SB. HS and AB helped SB in experiments. SB analyzed the data and wrote manuscript with help from AT. AT supervised the project. All authors contributed to this work and commented on this paper.

**Additional Information**

**Competing financial interests:**

The authors declare no competing financial interests.

**ACKNOWLEDGMENT**

We would like to acknowledge the support of Centre of Excellence in Nanoelectronics (CEN) at IIT-Bombay Nanofabrication facility (IITBNF), Indian Institute of Technology Bombay, Mumbai, India.

# Supplementary Material

Swapnil Bhuktare,[1] Hanuman Singh,[1] Arnab Bose,[1] and Ashwin. A. Tulapurkar[1,*]

[1]*Department of Electrical Engineering, Indian Institute of Technology-Bombay, Powai, Mumbai 400076, India*

**1) $S_{11}$ signal of device 1**

The $S_{11}$ spectrum as a function of frequency is shown below in Fig. S1. $S_{11}$ shows a dip at approximately 1.92 GHz, indicating surface acoustic waves are launched by IDTs at this particular frequency.

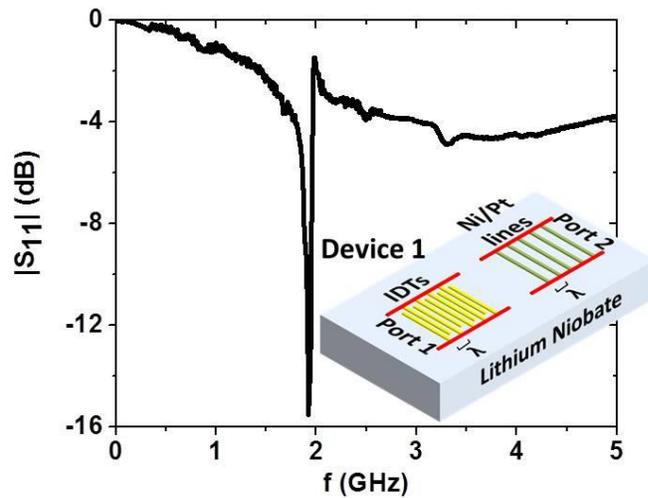

Fig. S1 $S_{11}$ spectrum of Device 1 showing a dip at ~1.92 GHz frequency..

**2) DC voltage spectra measured by ST-FMR**

ST-FMR measurements were carried out on Ni/Pt (device 2) and Ni/Au (device 3) samples. The dc voltage measured as a function of magnetic field applied at 45º angle, with rf current of 5 GHz are shown in Fig. S2(a) and (b) respectively. For Ni/Pt sample, the shape is almost like a peak showing spin current excitation and for Ni/Au sample, the shape is almost dispersion showing Oersted field excitation. The signal is small for Ni/Au sample as it is gets shorted by the highly conducting and much thicker Au.

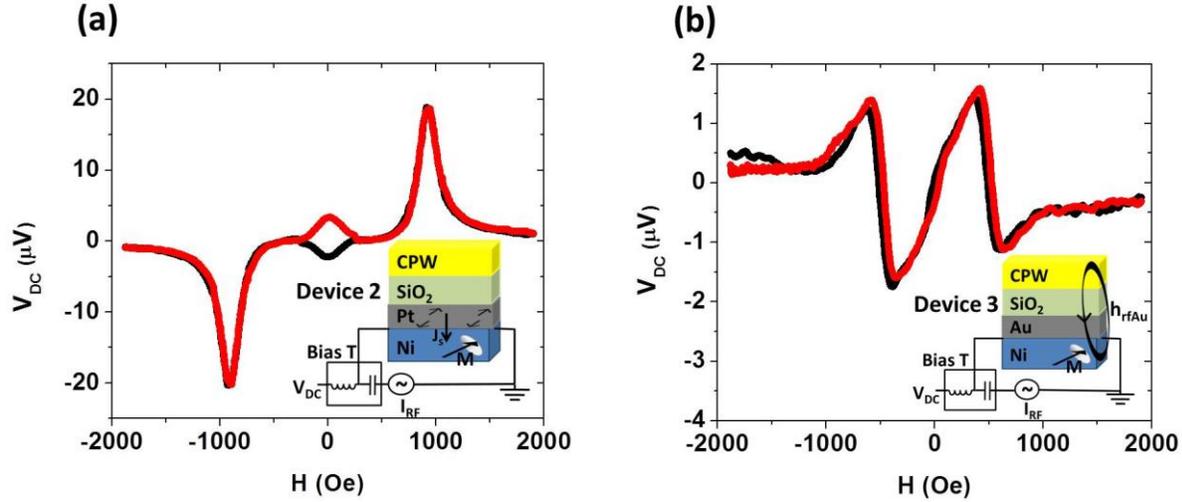

Fig. S2 The dc voltage measured with bias T as a function of magnetic field. The rf current frequency is 5 GHz, applied power is 10 dBm. The magnetic field was applied at an angle of 45 degree. (a) for Ni/Pt device (b) for Ni/Au device.

**3) S-parameters of device 2 and 3**

The $S_{12}$ and $S_{21}$ parameters of devices 2 (Pt/Ni) and 3 (Au/Ni) are shown in the paper. Here we show all S-parameters including $S_{11}$ and $S_{22}$ in Fig. S3 and Fig. S4 respectively. The left hand panel shows the raw S-parameters measured, while the right hand panel shows corrected ones. The raw S-parameters were measured after standard two port calibration of network analyzer using a calibration substrate. The $S_{11}$ and $S_{22}$ parameters are corrected as follows:

We find out the impedance of the Pt/Ni (or Au/Ni in case of device 3) away from resonance field (H=2000 Oe) and then correct the $S_{11}$ and $S_{22}$ parameters as follows

$$Z_1 = 50 \bullet \frac{(1+S_{11,}(2kOe))}{(1-S_{11}(2kOe))}, \quad \Delta S_{11}(H) = (S_{11}(H) - S_{11}(2kOe))\frac{(Z_1+50)^2}{|(Z_1+50)^2|}$$

$$Z_2 = 50 \bullet \frac{(1+S_{22,}(2kOe))}{(1-S_{22}(2kOe))}, \quad \Delta S_{22}(H) = (S_{22}(H) - S_{22}(2kOe))\frac{(Z_2+50)^2}{|(Z_2+50)^2|}$$

$Z_1$ and $Z_2$ are the impedances of the Ni/Pt and Au lines in case of device 2. Such a correction is required as these lines show inductive behavior. The factor of 50 Ω in above equations is the characteristic impedance of the cables.

The $S_{12}$ and $S_{21}$ parameters were corrected as follows:

$$Z_1(H) = 50 \bullet \frac{(1+S_{11}(H))}{(1-S_{11}(H))} \quad \text{and} \quad Z_2(H) = 50 \bullet \frac{(1+S_{22}(H))}{(1-S_{22}(H))}$$

$$S_{12,corrected}(H) = S_{12}(H) \frac{(Z_1(H)+50)(Z_2(H)+50)}{|(Z_1(H)+50)(Z_2(H)+50)|}$$

$$S_{21,corrected}(H) = S_{21}(H) \frac{(Z_1(H)+50)(Z_2(H)+50)}{|(Z_1(H)+50)(Z_2(H)+50)|}$$

The raw and corrected S parameters for Ni/Pt and Ni/Au samples are shown in Fig. S3 and Fig. S4 respectively.

**Comments on S-parameters**: The $S_{11}$ and $S_{22}$ parameters of both devices 2 and 3 have same features viz. $S_{11}$ and $S_{22}$ parameters are even functions of B i.e. $S_{11}(B)=S_{11}(-B)$ and $S_{22}(B)=S_{22}(-B)$. The real part shows a peak at resonance and the imaginary part shows dispersion behavior. Thus simply looking at $S_{22}$ data one can not make out the mechanism of resonance whether it is from the Oersted field or spin current. In the case of device 2, when we apply voltage to Ni/Pt line, the $S_{22}$ signal originates as follows: The magnetization of Ni is excited by spin current generated by SHE in Pt. Ni pumps back spin current into Pt which is converted into voltage by ISHE. This gives rise to the reflected voltage signal which is measured as $S_{22}$. The $S_{22}$ signal of device 3 originates as follows: The magnetization of Ni is excited by Oersted field generated by current flowing in Au (Ampere's law). A back emf is induced in Au according to Faraday's law. The $S_{11}$ signals in both the devices 2 and 3 also originate from the same mechanism.

When we measure $S_{22}$ of device 3, the Ampere's law and Faraday's law (and Lenz law) make sure that the real part is positive which correspond to positive dissipation. In the case of $S_{22}$ measurement of device 2, combination of STT and spin-pumping is such that $S_{22}$ is positive.

The $S_{12}$ (and $S_{21}$) parameter of the two devices is distinctly different. Real part shows peak and imaginary part shows dispersion in case of device 3. While in the case of device 2, real part shows dispersion and imaginary part shows peak/dip. This happens because the excitation is by spin current but the detection is by inductive coupling.

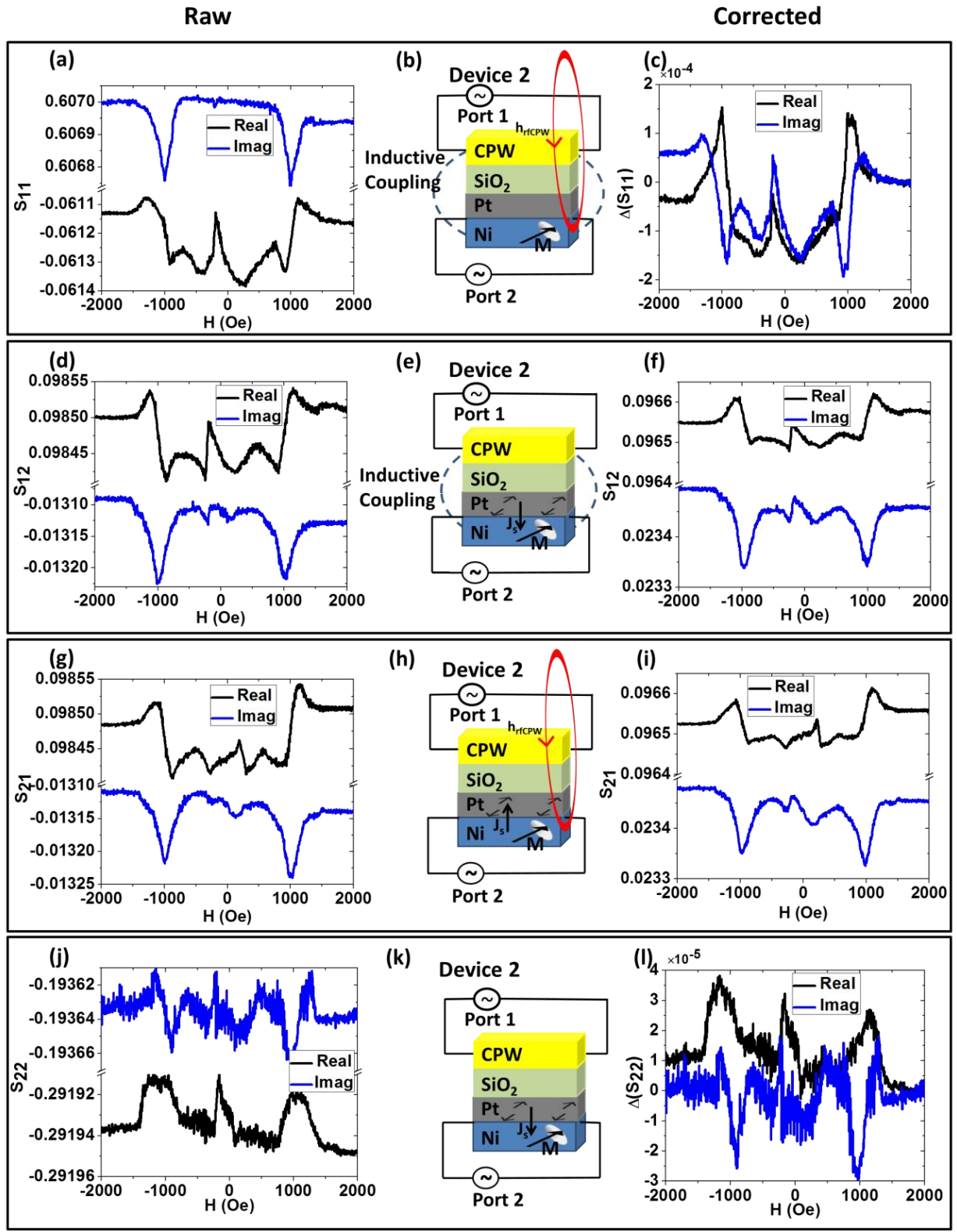

Fig. S3 The raw and corrected signals for Ni/Pt device. The left column shows raw signals, the middle column of figures shows how the particular signal is originating and the right column

show the corrected signals. (a), (b), (c) $S_{11}$ signal. (d), (e), (f) $S_{12}$ signal. (g), (h), (i) $S_{21}$ signal. (j), (k), (l) $S_{22}$ signal.

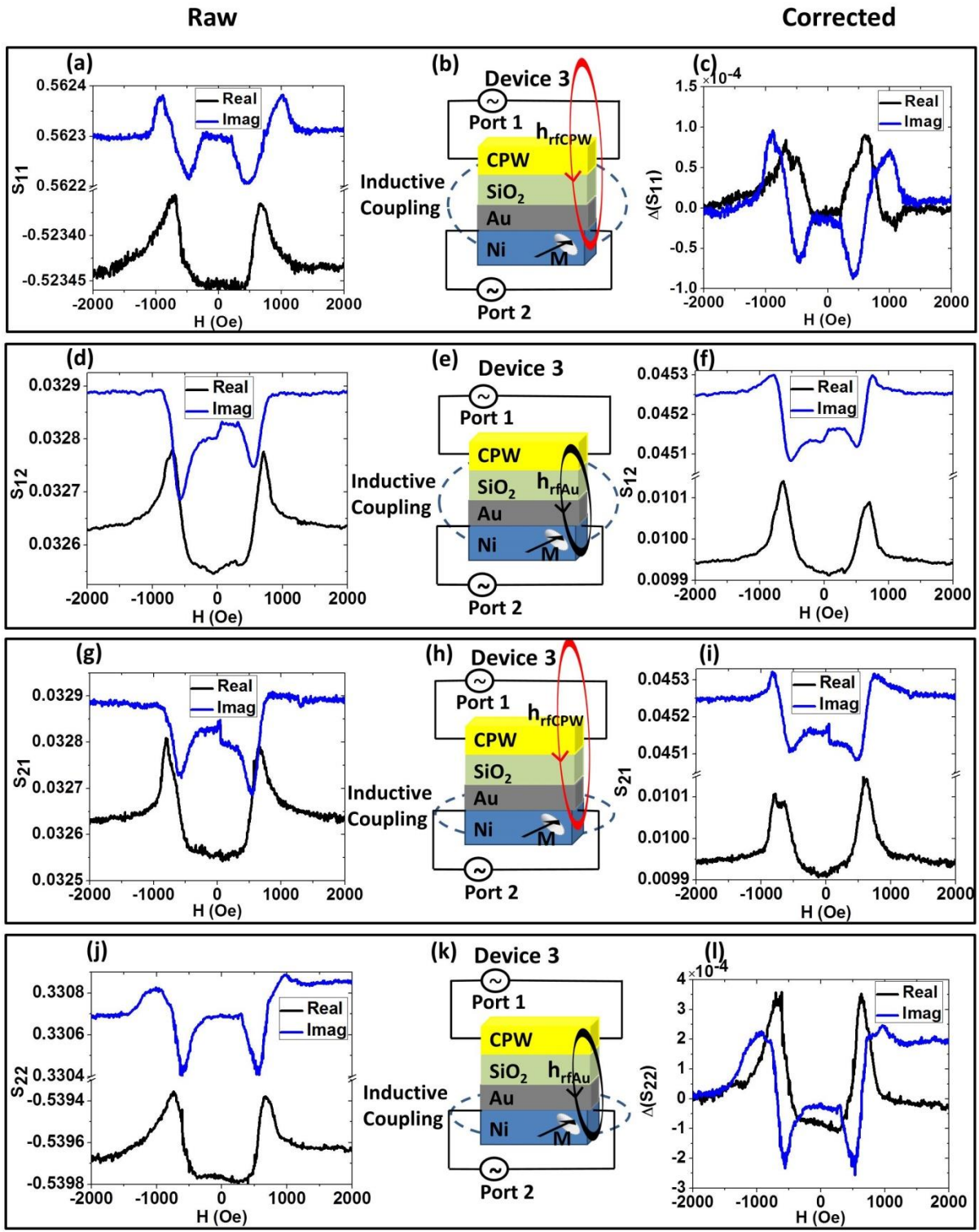

Fig. S4 The raw and corrected signals for Ni/Au device. The left column shows raw signals, the middle column of figures shows how the particular signal is originating and the right column show the corrected signals. (a), (b), (c) $S_{11}$ signal. (d), (e), (f) $S_{12}$ signal. (g), (h), (i) $S_{21}$ signal. (j), (k), (l) $S_{22}$ signal.

**4) Reciprocity between spin Hall and inverse spin Hall effect**

We have used the S-parameter measurements for a direct experimental proof of reciprocity between STT and spin pumping effects. The reciprocity between two effects can also be directly demonstrated by constructing a multi-terminal device and measuring resistance with interchange of current source and voltmeter connections, along with reversing magnetic field [S1] as shown in Fig. S5. Reciprocity between SHE and ISHE has been shown previously by using non-local methods [S2]. We here show it by using a local method. The three terminal device shown in two different configurations can be used for this purpose. This however gives rise to a large background signal. We therefore made a 5 terminal device and measured voltage in two different configurations as shown in Fig. S6(a) and (c) below. The device was fabricated with standard optical lithography, deposition and lift off processes. The voltage measured as a function of magnetic field after background subtraction is shown in Fig. S6. The voltage step indicated in Fig. S6(b) is due to the SHE, whereas the voltage step in Fig. S6(d) is due to the ISHE, which shows the reciprocity between SHE and ISHE. The large signal (peaks or dips) observed near H=0 is due to the magnetic domains and AMR effect of Ni, and some asymmetry in the experimentally fabricated device.

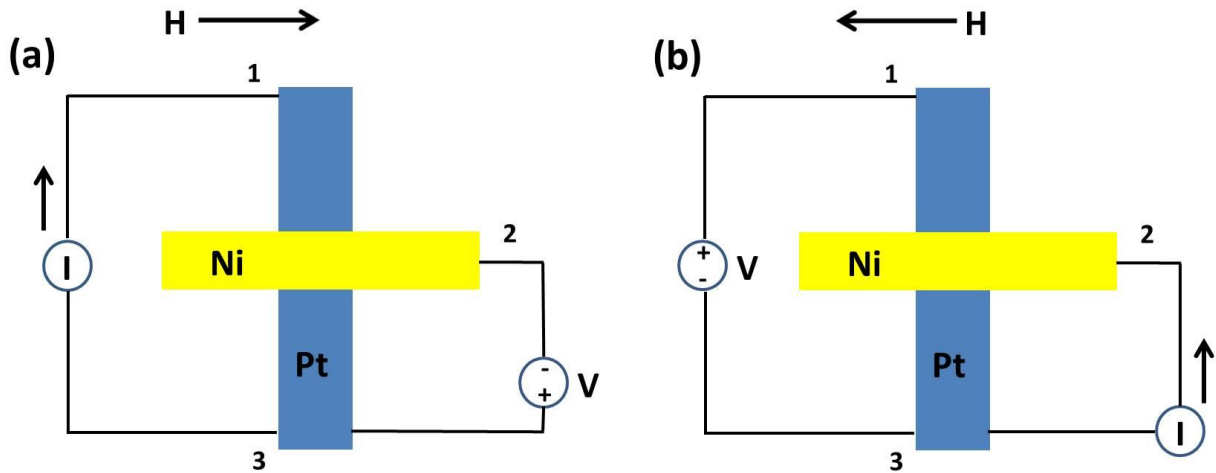

Fig. S5 (a) The schematic of the measurement setup for direct spin Hall effect measurement. Current is passed between terminals 1 and 3, and voltage is measured between terminals 3 and 2. (b) The schematic of the measurement setup for inverse spin Hall effect measurement. Current is passed between terminals 2 and 3, and voltage is measured between terminals 1 and 3.

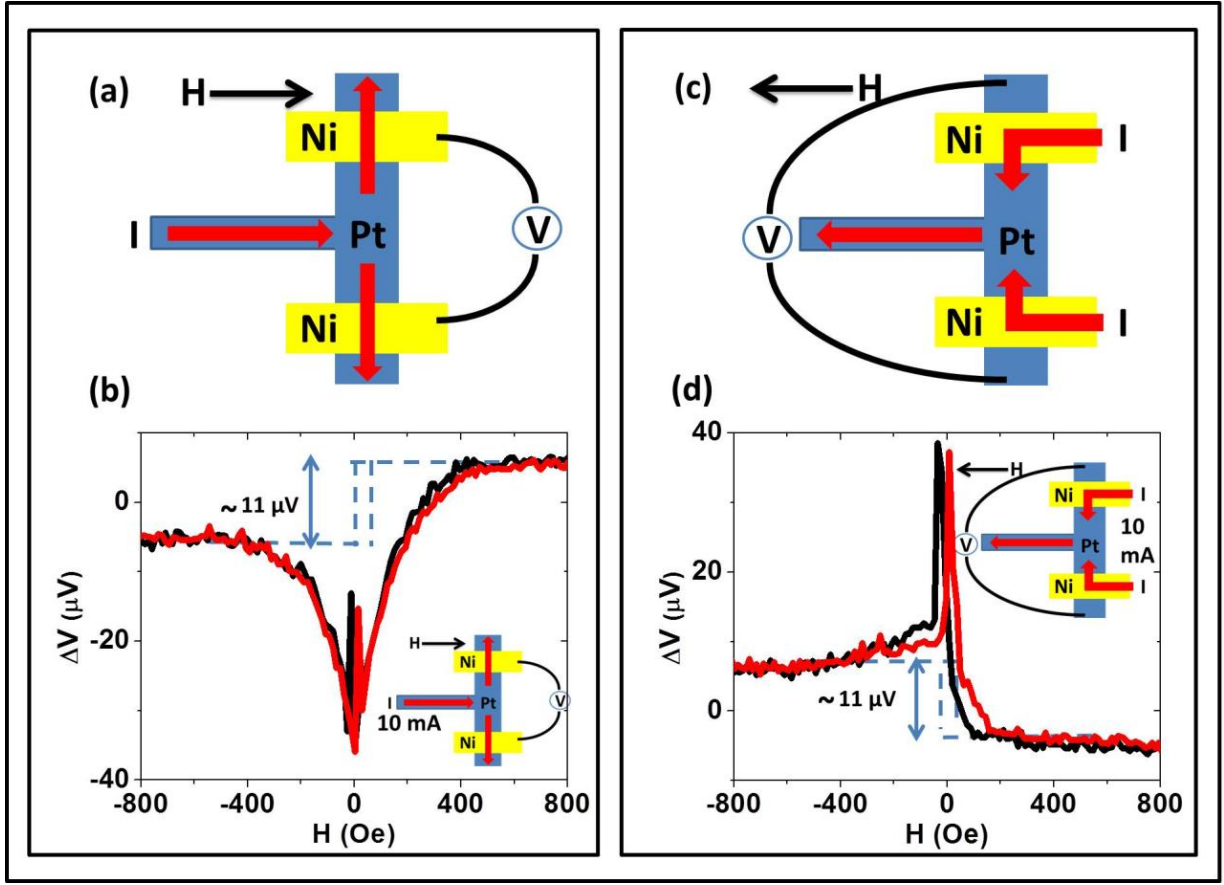

Fig. S6 (a) The schematic of the measurement setup for direct spin Hall effect measurement. (b) The direct spin Hall voltage signal for 10 mA current (c) The schematic of the measurement setup for inverse spin Hall effect measurement. (d) The inverse spin Hall voltage signal for 10 mA current

## 5) Reciprocity between STT and spin-pumping

The schematic of the device used for demonstrating reciprocity between STT and spin-pumping effect is shown below along with the reference frame used in Fig. S7.

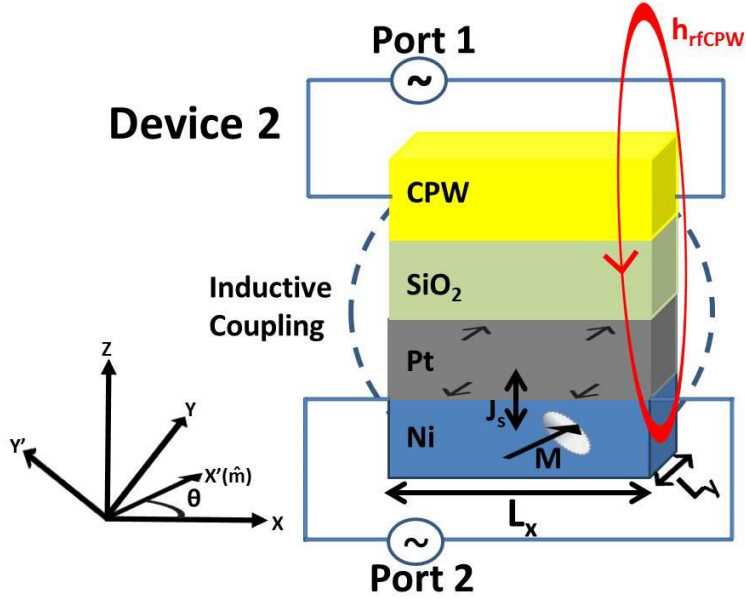

Fig. S7 The device schematic used to show the reciprocity between STT and spin-pumping

We assume that the equilibrium magnetization is in x-y plane, along x' axis which makes an angle θ with x-axis. The free energy density of the FM is taken as: $(\mu_0 M_s)[-\hat{m}\cdot\overline{H}_{ext} - (1/2)H_{//}m_x^2 + (1/2)H_\perp m_z^2]$. $\hat{m}$ denotes unit vector along magnetization direction and $M_s$ denotes saturation magnetization of FM. Thus positive values of $H_{//}$ and $H_\perp$ correspond to x axis as easy axis, and z axis as out of plane hard axis. The magnetic field acting on the FM is given by: $\overline{H} = -(1/\mu_0 M_s)(\partial F/\partial \hat{m}) = \overline{H}_{ext} + H_{//}m_x \hat{x} - H_\perp m_z \hat{z}$. The magnetization dynamics is governed by LLG equation as: $\partial_t \hat{m} = -\gamma_0(\hat{m}\times\overline{H}) + \alpha(\hat{m}\times\partial_t\hat{m}) + A_I \hat{m}\times(\overline{I}_S \times \hat{m})$, where $\gamma_0$ denotes the gyromagnetic ratio, α denotes damping factor and $I_S$ denotes the spin current in the units of charge current. The factor $A_I$ is given by, $A_I = \mu_B/eM_s vol$, where vol is the volume of FM and e is magnitude of electronic charge. Though, in the actual experiment, we have used Ni as a ferromagnet, we will assume here an insulating ferromagnet to simply the analysis. (e.g. an advantage is that when voltage is applied to port 2, we can assume that current flows only in the heavy metal.) We will now find out the transmitting and receiving characteristics of the ports 1 and 2, to calculate the $S_{12}$ and $S_{21}$ parameters assuming excitations with time dependence exp(-iωt).

We now consider transmitting characteristics of port 1. When port 1 is excited by voltage ($V_{in}$) from network analyzer, the current flowing in the Au line is given by, $I = 2V_{in}/(R_L + Z_1)$, where $Z_1$ includes the wire resistance and magnetic impedance. This current produces a magnetic field along y direction ($h_y = I/2L_y$) and excites the magnetization as:

$hy' = \cos\theta\, h_y$, $m_z = \chi^H_{21} h_{y'} = \chi^H_{21} \cos\theta\, h_y$

$$\Rightarrow m_z = \cos\theta\, \chi^H_{21} \frac{V_{in}}{L_y (R_L + Z_1)} \quad --- \quad (S1)$$

where $\chi^H$ denotes the susceptibility to magnetic field.

Now consider the receiving characteristics of port 1. The flux in the port 1 circuit from the magnetization of FM line is, $\varphi = (1/2)\mu_0 M_s L_x t_{FM}\, \delta m_y$, where $t_{FM}$ is the thickness of FM [S3]. If the magnetization is oscillating, it induces an emf in the circuit. The voltage measured by network analyzer is given by,

$$V_{out} = \frac{-\partial \varphi}{\partial t} \frac{R_L}{R_L + Z_2} = i\omega \frac{1}{2} \mu_0 M_s L_x t_{FM} \frac{R_L}{R_L + Z_1} \delta m_y \quad --- \quad (S2)$$

We now consider transmitting characteristics of port 2. When port 2 is excited by voltage ($V_{in}$) from network analyzer, the current flowing in the HM is given by, $I = 2V_{in}/(R_L + Z_2)$, where $Z_2$ includes the wire resistance and magnetic impedance. This corresponds to a charge current density, $J$ of $I/(L_y * t_{HM})$. We are going to ignore the magnetic field produced by this current and consider only the spin current produced by it via the spin-Hall effect. The spin current incident on the FM is given by equation (14) in ref. [S4]. We are going to assume that mixing conductance G is real and $G<<(\sigma/\lambda)$, where $\sigma$ is the conductivity and $\lambda$ is the spin diffusion length of HM. We further assume that the thickness of the HM metal $t_{HM}<<\lambda$. Under these assumptions, the spin current density is given by: $J_s = J\theta_{sh}(t_{HM}/\sigma_{HM})G_{\uparrow\downarrow}\,\hat{y}$. The spin current is given by $I_s = I\theta_{sh}(L_x/\sigma_{HM})G_{\uparrow\downarrow}\,\hat{y}$ and excites the magnetization as:

$I_{s,y'} = \cos\theta\, I_{s,y}$, $m_{y'} = \chi^{SC}_{11} h_{y'}$, $\delta m_y = \cos\theta\, m_{y'} = \cos^2\theta\, \chi^{SC}_{11}(A_I/\gamma) I_{s,y}$

$$\Rightarrow \delta m_y = \cos^2\theta\, \chi^{SC}_{11}(A_I/\gamma)\theta_{sh}(L_x/\sigma_{HM})G_{\uparrow\downarrow} \frac{2V_{in}}{(R_L + Z_2)} \quad --- \quad (S3)$$

where $\chi^{SC}$ denotes the susceptibility to spin current.

Now consider the receiving characteristics of port 2. When the magnetization of FM oscillates, it pumps spin current density into the HM given by, $\bar{J}_s' = (\hbar/4\pi)\text{Re}(g_{\uparrow\downarrow})(\hat{m}\times\partial_t\hat{m})$ [S5], where we have neglected the imaginary part of mixing conductance. The g and G are related as $G_{\uparrow\downarrow} = (2e^2/h)g_{\uparrow\downarrow}$. Further above expression for spin current density converted into units of charge current density reads as, $\bar{J}_s = (e/2\pi)\text{Re}(g_{\uparrow\downarrow})(\hat{m}\times\partial_t\hat{m})$. The y-component of spin current density induces charge current density along x-direction via inverse spin-Hall effect. The voltage measured by network analyzer is found out using the following steps:

$$J_x = \theta_{sh} J_{s,y}, E_x = J_x / \sigma_{HM}, V_{oc} = E_x L_x$$

$$V_{out} = V_{oc} \frac{R_L}{R_L + Z_2}$$

$$use: [\hat{m} \times \frac{d\hat{m}}{dt}]_y = i\omega \cos\theta \delta m_z$$

$$\Rightarrow V_{out} = \frac{e}{2\pi} g_{\uparrow\downarrow} i\omega \cos\theta \, \theta_{sh} \frac{L_x}{\sigma_{HM}} \frac{R_L}{R_L + Z_2} \delta m_z \quad ---(S4)$$

The parameter $S_{12}$, which corresponds to excitation at port 2 and measurement at port 1 is can be found from equations (S3) and (S2) as ($S_{12}=V_{out}/V_{in}$):

$$S_{12} = i\omega \, \cos^2\theta \, \chi_{11}^{SC} \frac{\hbar}{2e} \theta_{sh} \frac{R_L}{(R_L+Z_1)(R_L+Z_2)} \frac{L_x}{L_y} \frac{G_{\uparrow\downarrow}}{\sigma_{HM}}$$

where we have used $(A_I/\gamma) = \hbar/2eM_s\mu_0 vol$.

Similarly parameter $S_{21}$, which corresponds to excitation at port 1 and measurement at port 2 is can be found from equations (S1) and (S4) as ($S_{21}=V_{out}/V_{in}$):

$$S_{21} = i\omega \, \cos^2\theta \, \chi_{21}^{H} \frac{\hbar}{2e} \theta_{sh} \frac{R_L}{(R_L+Z_1)(R_L+Z_2)} \frac{L_x}{L_y} \frac{G_{\uparrow\downarrow}}{\sigma_{HM}}$$

where we have used, $G_{\uparrow\downarrow} = (2e^2/h)g_{\uparrow\downarrow}$. Using the fact that, $\chi_{21}^{H} = \chi_{11}^{sc}$, we see that $S_{12}=S_{21}$.

### 6) Numerical estimation of $S_{21}$ signal at resonance

When SAW in incident on Ni/Pt line, it exerts an effective magnetic field. The value of magnetic field has been estimated in ref [S6]. (See supplementary information. The parameters used here and ref [S6] are almost the same). For a 5 dBm of input power at IDT, we get an effective magnetic field of 1.19 Oe. When the magnetization of Ni undergoes resonance, it pumps spin current into Pt, which induces voltage via ISHE. Once we know the oscillating magnetic field, we can use the equations leading to derivation equation (S4) to find out voltage measured at port 2. We have used the following material parameters to calculate the induced voltage: $\omega=2\pi*1.92 \times 10^9$, Ms=800 emu/cc, $\alpha$=0.05, W=300 μm, t=15 nm, $R_L$=50 Ω, $Z_2$=30 Ω and $H_{ext}$=100 Oe (which corresponds to resonance), $g_{\uparrow\downarrow} = 2x10^{19}/m^2$, $\sigma_{Pt} = 10^7/ohm-m$. We can then get $S_{12}$ as ratio of induced voltage to applied voltage comes out to be, $|S_{21}|=8 \times 10^{-6}$, which is comparable to the observed value (see Fig. S8).

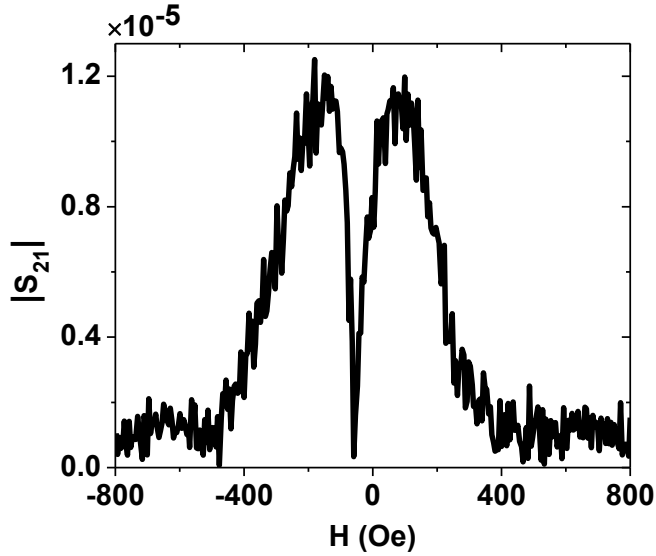

Fig. S8 The magnitude of the $S_{21}$ signal

### 7) S parameters for Device 1 with less Ni thickness

The S parameters for device 1 with Ni (4 nm )/Pt (8 nm) are shown in Fig. S9

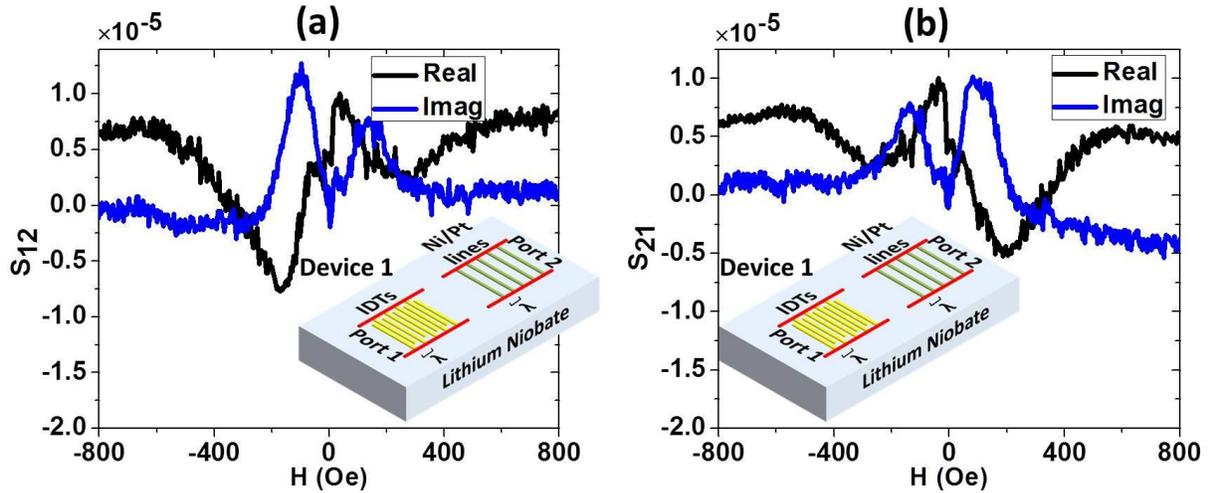

Fig. S9 Transmission signals for Device 1 with less Ni thickness (4 nm) (a) $S_{12}$ signal (b) $S_{21}$ signal

We see that this device satisfies the generalized reciprocity relation but does not show gyrator behavior observed for device with Ni (8 nm)/Pt (6 nm). The $S_{12}$ (and $S_{21}$) signal shown in Fig. S9 is composed of both symmetric and anti-symmetric parts. The anti-symmetric part originates from magneto-elastic coupling. The symmetric part could arise from the electric field associated with SAW. This electric field can induce oscillatory current and give rise to resonance [S7]. We observe that as we increase the metal thickness, the effect of the electric field is reduced and we can clearly see a gyrator behavior as shown in the main paper in Fig. 3. The observation of gyrator behavior itself proves that magneto-elastic coupling between the ferromagnet and the piezoelectric substrate dominates over electric field effect [S6].